\documentclass[aps,prl,twocolumn,groupedaddress,showpacs]{revtex4}

\usepackage{graphicx}
\usepackage{epstopdf} 
\DeclareGraphicsExtensions{.pdf,.eps,.png,.jpg,.mps} 
\usepackage{amssymb}
\usepackage{color}

\newcommand{\rs}{\rm \scriptscriptstyle}

\begin{document}

\title {Supersolid phase in atomic gases with magnetic dipole interaction}

\author{Adam B\"uhler and Hans Peter B\"uchler}
\affiliation{Institute for Theoretical Physics III, University of Stuttgart, Germany}

\date{\today}

\begin{abstract}
A major obstacle for the experimental realization of a supersolid phase with cold atomic gases
 in an optical lattice is the weakness of the nearest-neighbor 
interactions achievable via magnetic dipole-dipole interactions.
 In this letter, we show that using a large filling of atoms within each well 
the characteristic energy scales are strongly enhanced. Within this regime, the system is well described
by the rotor model, and the qualitative behavior of the phase diagram derives from 
 mean-field theory. We find a stable supersolid phase for realistic parameters with
chromium atoms. 
\end{abstract}


\maketitle

Magnetic dipole-dipole interactions offer a remarkable opportunity to explore
quantum phenomena with long range interactions in cold atomic gases.  Of
special interest are atoms with large magnetic dipole moments such as Cr, where
the influence of the dipole-dipole interactions on the atomic cloud as well as
the dipole induced collapse have been observed \cite{Pfau_2005_PhysRevLett.94.160401,NaturePhys_Pfau_4,Beaufils_PRA_2008_PhysRevA.77.061601}.  
In the presence of an optical lattice, the system gives naturally rise to extended Hubbard models
with nearest-neighbor interactions \cite{Goral_2002_PhysRevLett.88.170406}, and many remarkable
quantum states have been predicted 
\cite{Schmid_2005_PhysRevLett.94.207202,Pupillo_2010_PhysRevLett.104.125301,Pollet_2010_PhysRevLett.104.125302,Danshita_2009_PhysRevLett.103.225301}. A major obstacle towards 
the experimental realization of these states is
the weakness of the nearest-neighbor interaction due to the magnetic character
of the dipole-dipole potential, and correspondingly the extremely stringent
requirements on temperature, trapping potentials, and life-time of the atomic
system.  In this letter, we propose an experimentally realistic setup for the
realization of a supersolid phase in cold atomic gases with magnetic
dipole-dipole interactions.

A supersolid phase combines two seemingly contradiction properties, which in
most materials appear in competition with each other: the arrangement of the
particles in a crystalline structure with a superfluid transport of the
particles \cite{Leggett_1970_PhysRevLett.25.1543}. While recent experimental observation of a
superfluid response in solid $^{4}{\rm He}$ are still controversial \cite{Nature_Kim_2004,Reppy_2010_PhysRevLett.104.255301,Clark_2006_PhysRevLett.96.105302},
various models in lattice systems have extensively been studied in the past and
the existence of a supersolid phase has been demonstrated using quantum Monte
Carlo simulations. Of special interest is the appearance of a supersolid phase
in a triangular lattice \cite{Wessel_2005_PhysRevLett.95.127205} and the stabilization of a supersolid phase in
a square lattice by  dipole-dipole interaction \cite{Pupillo_2010_PhysRevLett.104.125301}. These  Hubbard models
can be naturally realized with cold polar molecules or atomic gases with
magnetic dipole-dipole interactions. While the understanding of the microscopic
derivation of the Hubbard model for cold atomic gases is well understood \cite{Jaksch_1998_PhysRevLett.81.3108}, the
nearest-neighbor interactions obtained for atoms in a characteristic optical
lattice is well below a nano Kelvin, i.e., for Cr with strong magnetic
dipole moment $V\approx 0.2 {\rm nK}$  with lattice spacing $a = 500 {\rm nm}$ -- 
such stringent requirements have not yet been achieved in experiments.

In this letter, we demonstrate that the critical temperature for the supersolid 
phase can be increased by several orders of magnitude. This opens up a way
towards the experimental realization of supersolids with cold atomic gases using 
the magnetic dipole-dipole interactions with current experimental technologies.
The main idea is to allow a high filling factor of atoms per lattice site; then the influence of
the dipole-dipole interaction is enhanced by the number of atoms within each well. In the extreme
situation, the system is then described by a coupled array of Bose-Einstein condensates.

\begin{figure}[ht]
 \includegraphics[width= 0.9\columnwidth]{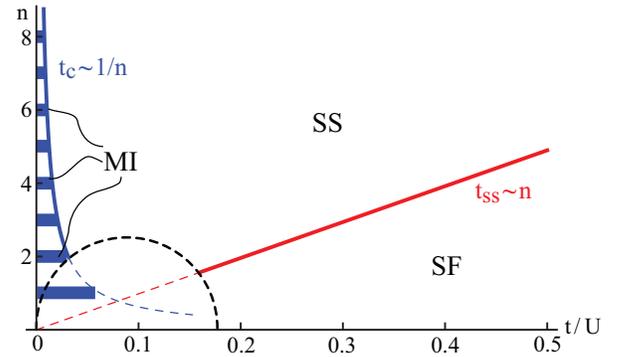}
  \caption{Phase diagram for $U \!\sim\! V$: The Mott insulating/solid phases appear at commensurate fillings (solid bars) and the critical hopping at the tip of the incompressible lobes scales as  $t_{c} \sim 1/n$. In turn, the transition from the superfluid into a supersolid phase behaves as $t_{\rs ss} \sim n$, i.e., for large fillings $n\gg 1$ the influence of quantum fluctuations is reduced and the transition from the superfluid into the supersolid phase is well described within mean-field theory.}   \label{Fig1}
\end{figure}

For low filling factors, the competition between the instability towards a
supersolid phase and the Mott insulating phase for integer fillings gives rise
to a rich phase diagram. Within this regime mean-field theory fails to
correctly reproduce the phase diagram and it is mandatory to resolve to exact
methods such as quantum Monte Carlo simulations \cite{Hebert_2001_PhysRevB.65.014513}. 
On the other hand, for high
filling factors the critical hopping $t_{c}$ for the phase transition towards
the solid phases/Mott insulators scales as $t_{c} \sim 1/n$, while the
instability from the superfluid towards the supersolid phase appears at much
higher hopping with $t_{\rm ss} \sim n$, see Fig.~\ref{Fig1}.  Consequently,
the competition between the different phases is reduced and the phase diagram
is well described within mean-field theory. In the following, we derive the
transition towards the supersolid phase  and find the
optimal parameters for the experimental realization in cold atomic gases with
magnetic dipole-dipole interactions.

We start with the description  of  the Hamiltonian for the cold atomic gases
with magnetic dipole-dipole interactions.  The system is confined into
a quasi two-dimensional setup with an additional optical lattice within the
plane.  Each lattice site is occupied by many particles, and therefore, gives
rise to a quasi-condensate on each lattice site. Note,  that we assume a lattice spacing larger than the usual
spacing in optical lattices in order to reduce losses from  three-body recombinations.
For such high filling factors, the validity of the standard Hubbard model breaks down.
Nevertheless, the system exhibits tunneling between the different wells,  and
an interaction term accounting for deviations around the mean particle number $n$.
Then, the Hamiltonian is well described by the rotor model
\begin{equation}
 	H = -2t \sum_{\langle i j \rangle} \sqrt{ n_{i}  n_{j}} 
	\cos\left(\phi_{i}\!-\!\phi_{j}\right) + \frac{1}{2} \sum_{i j} V_{i j} \delta n_{i}  \delta n_{j} \, ,
 	\label{eq:HRotor001}
 \end{equation}
where $\delta n_{i} = n_{i} - n$ describes the deviation from the mean particle
density $n$ within each well, while $\phi_{i}$ denotes the phase within each
well satisfying the commutation relation $\left[n_{i}, \phi_{j}\right] = i \delta_{i j}$.
The interaction term $V_{i j}$ contains an on-site interaction $U$ for $i=j$, and
obeys the characteristic decay $V_{i \neq j} = V  a^3/|{\bf R}_{i}-{\bf R}_{j}|^{3}$ for dipole-dipole interactions
with $a$ the lattice spacing and ${\bf R}_{i}$ the lattice vectors.
Note, that the rotor model derives from the Hubbard model in the limit of large filling factors,
however, the rotor model remains a proper description of bosonic atoms in an optical lattice
even in the regime where several higher bands are occupied.  Its phase diagram has previously been 
studied close to half filling $n\sim 1/2$  \cite{Wagenblast_1994_PhysRevLett.72.3598,Roddick_1995_PhysRevB.51.8672}.
Here, we first derive the phase diagram of the rotor model at large filling $n \gg 1$, and present the effective parameters for a realistic experiment with chromium atoms
in a second step.

The system is in the superfluid phase for dominant hopping with $t \gg V, U$, and is characterized 
by a fixed phase  $\phi$ within each well  and a homogeneous particle density $n$.
Its mean-field energy per lattice site reduces to  $E_{0}/N =  (U + V \chi_{0} ) n^2/2 - 2 t n z$ 
with $z=4$ the number of nearest-neighbors and $N$ the number of lattice sites.  Here,  
$\chi_{\bf k} = \sum_{j\neq 0} \exp(i {\bf k} {\bf R}_{j})/|{\bf R}_{j}|^{3}$ denotes the dipole-dipole 
interaction in momentum space. The transition towards the supersolid phase for increasing 
interactions is signaled by an instability in the excitation spectrum. Expanding the Hamiltonian to 
second order in  fluctuating fields $\delta n_{i} = n_{i}-n$  and $\delta \phi_{i} = \phi_{i}-\phi$ 
around the mean-field values, and introducing the momentum representation
with $\delta \phi_{\bf k} = \sum_{j} \exp(i {\bf k} {\bf R}_{j}) \delta \phi_{j}/\sqrt{N} $ and  analog for 
$\delta n_{\bf k}$, we obtain the Hamiltonian $H_{\rs fl}$ describing the excitation spectrum
above the superfluid ground state
\begin{displaymath}
	H_{\rs fl} = \sum_{{\bf k}} \left(  \epsilon_{{\bf k}} \, \delta \phi_{{\bf k}} \delta \phi_{-{\bf k}} 
	+ \widetilde{V}_{{\bf k}}  \delta n_{{\bf k}} \delta n_{-{\bf k}} \right) =  \sum_{\bf k} E_{\bf k} a^{\dagger}_{\bf k} a_{\bf k}  ,	\label{eq:HRotor003}
\end{displaymath}
with the single particle dispersion relation $\epsilon_{\bf k}  =  2 t n \left[ z - 2 \sum_{\alpha} \cos\left( {\bf k} {\bf e}_{\alpha}\right)\right] $ and the effective interaction 
$ \widetilde{V}_{\bf k}  =   \epsilon_{\bf k} /(2 n)^2+
	 U\left( 1 + \gamma \; \chi_{\bf k} \right)/2$.
Here,  $\gamma = V/U$ denotes the ratio between the strength of the dipole-dipole interaction $V$ 
and the on-site interaction $U$, whereas ${\bf e}_{\alpha}$ accounts for the unit vectors in direction of the 
nearest-neighbor lattice sites. Introducing the creation (annihilation) operators $a^{\dag}_{\bf k}
$ ($a_{\bf k}$) for the excitations above the superfluid ground state  through
$a^{\dag}_{{\bf k}}= i \beta_{\bf k}  \delta \phi_{\bf k} \!+ \!  \delta n_{\bf -k}/ 2\beta_{\bf k}$  with
 $\beta_{\bf k}^{4} =\epsilon_{\bf k}/4 \widetilde{V}_{{\bf k}}$,
we obtain the  excitation spectrum  of the superfluid phase $E_{\bf k} = (4 \widetilde{V}_{\bf k} \epsilon_{\bf k})^{1/2}$.

\begin{figure}[ht]
 \includegraphics[width= 1\columnwidth]{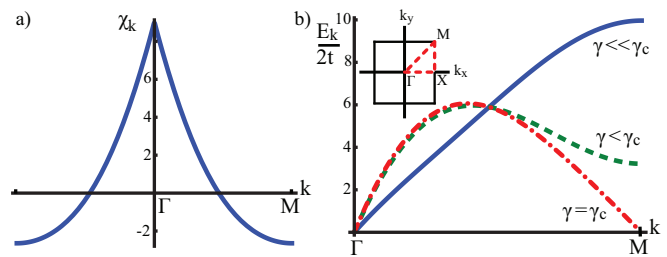}
  \caption{ \textbf{a)} Shape of the Fourier transformed dipole-dipole interaction $\chi_{\bf k}$ plotted in the $M$-direction. \textbf{b)} Superfluid dispersion relation for different values of $\gamma$. At $\gamma = \gamma_{c}$ the Roton minima occurs. The inset shows the different directions within the Brillouin zone.}   \label{Fig2}
\end{figure}

Next, we analyze the stability of the superfluid phase by varying the ratio $\gamma$ 
between the dipole strength and the on-site interaction.  The result strongly depends
on the lattice geometry. Here, we focus on a square lattice with 
lattice spacing $a$; the generalization to arbitrary lattice structures is straightforward.  
The quantity $\chi_{\bf k}$ is maximal at zero momentum with 
$\chi_{0}\approx9.02771$, while it turns negative and minimal at the edge of the Brillouin 
zone with  ${\bf K} = (\pi/a,\pi/a)$ and $\chi_{\bf K}\approx -2.64589 $, see Fig.~\ref{Fig2}a. As a consequence, 
the excitation spectrum exhibits a Roton minima for increasing dipole-dipole interactions, 
see Fig.~\ref{Fig2}b, and eventually turns zero $E_{\bf K}=0$ at the critical value
\begin{equation}
	\gamma_{c} = \frac{1}{\left| \chi_{\bf K} \right|} \left( \frac{2 z t}{U n} + 1 \right) .
		\label{eq:CritGamma}
\end{equation}
Hence, the superfluid phase suffers an instability at $\gamma_{c}$ via the nucleation of
excitations with momenta ${\bf K}$. In general, one expects that these excitations
form a second condensate and give rise to a density modulation for the
system. The novel ground state is characterized by a superfluid response due to 
the condensates at ${\bf k}=0$ and ${\bf K}$, and a solid order, i.e., the instability signals a 
phase transition  from  a superfluid into a supersolid.
 
Now, we analyze the stability and the ground state properties of the novel phase by 
a  mean-field ansatz with two condensates. For this purpose, we introduce a density 
modulation $\bar{n}_{j} = \left|c + d \exp\left[i \left( {\bf K} {\bf R}_{j}+\theta\right) \right] \right|^{2}$ 
with the constraint  $c^{2} + d^{2} = n$.  The density modulation appears as
the inference of the two condensates at ${\bf k}=0$ and ${\bf K}$ with the relative phase  $\theta$, and exhibits
a checkerboard structure. Inserting this ansatz
into the rotor Hamiltonian~(\ref{eq:HRotor001}), we obtain the 
energy  per lattice site 
\begin{equation}
	\frac{E_{\rs ss}(d,\theta)}{N}  =  \frac{ E_{0}}{N} +  4zt   \: d^{2} - 2d^2 \left(n -d^{2}\right)  \cos^2\theta  \left( \left| \chi_{\bf K} \right| V - U \right) . 
	\label{eq:HRotor007}
\end{equation}
Here, $E_{0}$ is again the mean-field energy of the superfluid phase. 
The second term accounts for an increase of kinetic energy due to a reduction of coherence  
between the different wells, while the last term describes the lowering of the interaction energy via the density 
modulation. The ground state is obtained by minimizing the energy with respect to the density modulation 
and the relative phase between the two phases. The optimization of the phase requires $\theta= 0,  \pi$, which 
corresponds to the two degenerate ground states reflecting the broken (discrete) translational symmetry. 
In the following, this phase will be absorbed into the sign of $d$.
On the other hand, the condensate fraction $d$ for the mode ${\bf K}$ exhibits the typical Ginzburg Landau
behavior for a second order phase transition and predicts the lowering of the ground state energy for 
$\gamma> \gamma_{c}$, i.e.,  the appearance of the supersolid phase within mean-field theory coincides with 
the instability of  the excitation spectrum. 
The gain in energy via the formation of the supersolid phase takes the form
\begin{equation}
	\frac{E_{\rs ss}- E_{0}}{N}= -\frac{ n^2 V}{2} \frac{|\chi_{\bf K}|^2 }{|\chi_{\bf K}|-1/\gamma} \left(1-\frac{\gamma_{c}}{\gamma}\right)^2 .
	\label{eq:HRotor009}
\end{equation}
A special property of the energy gain is, that it scales with the square of the number of 
particles per lattice site $n$. 
The supersolid phase is characterized by a checkerboard density modulation with
the order parameter $\Delta$ defined by the correlation function
\begin{equation}
    \langle \bar{n}_{i} \bar{n}_{j}\rangle =n^2 \left[ 1 + \Delta^{2} \cos\left({\bf K} \left( {\bf R}_{i} - {\bf R}_{j} \right) \right) \right] ,
\end{equation}
for $|i-j| \rightarrow \infty$ and a superfluid density $n_{s}$ describing the superfluid flow.
Within mean-field theory, these quantities reduce to 
$ n_{s} =  2 t z/(   |\chi_{\bf K}| V- U)$, and  $\Delta^2 = 1- n_{s}^2/n^2$. 
Note, that within a supersolid, the superfluid density is reduced compared to the averaged density, i.e.,  $n_{s}<n$ 
even at zero temperature due to the additional solid structure \cite{Leggett_1970_PhysRevLett.25.1543}.

The most remarkable result in Eq.~(\ref{eq:HRotor009}) is the scaling
of the gain in energy via
the formation of the supersolid phase  with the number of particles
in each lattice site,
i.e., $ (E_{\rs ss}- E_{0})/N \sim n^2 V/2$.   This energy serves as an
estimate for the critical temperature  $T_{\rs solid}$
for the formation of the solid structure, i.e., at optimal values the critical temperature
$T_{\rs solid} \sim n^2 V/2$ is strongly enhanced compared to the single particle 
nearest-neighbor energy $V$.  This
observation is in agreement with the recently predicted formation of
solid structures in very large superlattices \cite{Pfau_2010_PhysRevLett.104.170404}.
The detailed estimation of the energy scales for realistic
experimental setup is shown below.

\begin{figure}[ht]
 \includegraphics[width= 1\columnwidth]{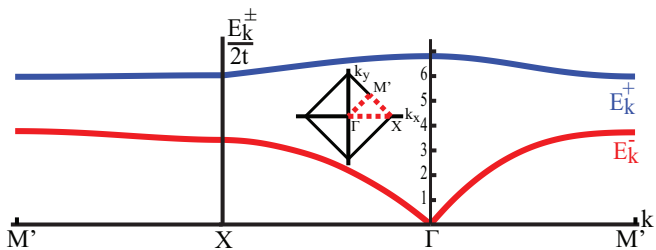}
  \caption{ Supersolid dispersion relation, given in different directions of the reduced Brillouin zone, as shown in the inset.}   \label{Fig3}
\end{figure}

Finally, we  check the stability of the supersolid phase against quantum fluctuations and 
determine the excitation spectrum above the
supersolid ground state. Again we introduce the fluctuating field operators $\delta \phi_{i} = \phi_{i}- \phi$
and $\delta n_{i} = n_{i}- \bar{n}_{i}$, where 
$\bar{n}_{i}/n= 1+\Delta \; \cos ( {\bf K} \: {\bf R}_{i} )$
denotes the mean particle density per lattice site within
the mean-field theory exhibiting a checkerboard structure. Then, the Hamiltonian expanded to second order
in these operators reduces to
\begin{eqnarray}
	H_{\rs ss} & =  &t n_{s} \sum_{\langle i, j \rangle} \left[  \left( \delta \phi_{i} - 
	   \delta \phi_{j} \right)^{2}  -   \frac{ \delta n_{i}  \delta n_{j} }{ 2 n_{s}^{2}}  \right]\\
	 & & \hspace{-30pt} +  \frac{t z}{2 n_{s}} \sum_{i} \delta n_{i}^2 \frac{ 1 + \Delta^{2} - 2 \Delta \cos\left( {\bf K} {\bf R}_{i} \right) }{ 1-\Delta^{2} } + \frac{1}{2} \sum_{i j} V_{i j} \delta n_{i} \delta n_{j} .  \nonumber
			\label{eq:HRotor01SS}
\end{eqnarray}
Note, that the third term involves the modulated density with wave vector ${\bf K}$.  As a consequence, 
this introduces a coupling for modes with momentum ${\bf k}$ and ${\bf k}+{\bf K}$, which becomes obvious in the moment
representation 
\begin{displaymath}
	H_{\rs ss}  = \sum_{k} \left[ \xi_{{\bf k}} \;\delta \phi_{{\bf k}}  \delta \phi_{-{\bf k}} + \zeta_{{\bf k}} \; \delta n_{{\bf k}}  \delta n_{-{\bf k}} + \eta \;  \delta n_{{\bf k}}  \delta n_{- ({\bf  k} + {\bf K}) } \right] ,
	\label{eq:HRotor02SS}
\end{displaymath}
with the  parameters $ \xi_{{\bf k}} =  2 t n_{s} [z - 2 \sum_{\alpha} \cos\left( {\bf k} {\bf e}_{\alpha}\right) ]$, 
\begin{displaymath}
	\zeta_{{\bf k}} =  \frac{ U  + \chi_{\bf k} V}{2} + \frac{z t}{2 n_{s}} \left[  \frac{ 1 + \Delta^{2} }{ 1 - \Delta^{2} } - \frac{2}{z} \sum_{\alpha} \cos\left( {\bf k} {\bf e}_{\alpha} \right) \right],  \nonumber \\
\end{displaymath}
and $\eta \, n_{s} = - z t \, \Delta /(1-\Delta^2)$.
This coupling is a result of the broken translational symmetry in the supersolid phase, which 
reduces the Brillouin zone. Hence, we obtain two modes for each momentum 
value ${\bf k}$ with the dispersion relation, see Fig.~\ref{Fig3},
\begin{eqnarray}
	\left(E^{\pm}_{\bf k}\right)^2 & = & 2 \left( \xi_{\bf k}  \zeta_{\bf k} +\xi_{\bf k + K}  \zeta_{\bf k + K} \right) \label{eq:HRotor03SS} \\  
			& \pm & 2 \sqrt{ \left( \xi_{\bf k}  \zeta_{\bf k} - \xi_{\bf k + K} \zeta_{\bf k + K} \right)^{2} +4  \xi_{\bf k}  \xi_{\bf k + K} \, \eta^{2}    } \nonumber . 
\end{eqnarray}
The lower branch of the dispersion relation accounts for the acoustic modes with a linear sound mode for small values ${\bf k}$, 
while the second branch accounts for density fluctuations of the checkerboard order. 
Due to the discrete translational symmetry this mode is lifted to a finite value for small values 
${\bf k}$ and corresponds to an optical mode. From the dispersion relation Eq.~(\ref{eq:HRotor03SS}), we find that
the supersolid phase is stable.

Finally, we can estimate the relevant parameters for the experimental
realization of a
supersolid phase for chromium atoms in an optical lattice
\cite{Pfau_2010_RPP}. The limiting energy is given by the nearest-neighbor 
interaction $V$ for magnetic dipole-dipole interactions.
These interactions are characterized by the length scale
$a_{dd} = \mu_{0} \mu^{2} m/(12 \pi \hbar^{2})   \approx   0.8  {\rm n m}$,
where $\mu$ is the permanent magnetic dipole moment. Then, the strength $V$ of the dipole-dipole 
interaction  reduces to
\begin{equation}
	V =  \frac{6}{\pi^{2}}    \frac{a_{dd}}{a}  E_{r} ,  	\label{eq:HRotor013}
\end{equation}
with $E_{r}= \pi^2 \hbar^2/2 m a^2$ the recoil energy and $a$ the lattice spacing. 
 Starting with a
conventional density for a BEC of
cold atomic gases, we obtain a filling with $n \sim 40$ for a lattice
spacing of $a \sim 1 \mu {\rm  m}$
within the plane and a prolate shape of each well along the
perpendicular direction with aspect
ration $\lambda \approx 1/4$.  The nearest-neighbor energy reduces to
$V \approx 0.5 {\rm Hz}$, while the characteristic temperature scale for
the formation
of the solid structure reaches $T_{\rs solid}\sim n^2 V/2 \approx 0.4 E_{r}
$. On the other hand, the on-site interaction
within each well derives as the change of energy for the local
condensate within the well
by adding/removing particles, i.e., $U = \partial_{n}^2E_{\rs
local}[n]$. Tuning the s-wave scattering
length $a_{s}$ across the Feshbach resonance present for chromium
allows us to compensate
the contribution from the dipole-dipole interaction, and consequently
the on-site interaction can be tuned
to an arbitrary small value. For particle numbers with $n\sim 40$ within
the quasi-condensate, the influence of the interactions is weak on the
ground state wave function, which
is dominated by the contribution from the trap and kinetic energy.
Then, the ground state wave function is well described
by a Gaussian wave function and the on-site interaction reduces to
(see Ref.~\cite{Pfau_2010_RPP} for details)
\begin{equation}
	U = \frac{2 \hbar \overline{\omega} a_{dd}}{\sqrt{2 \pi} a_{ho} }
	 \left[ \frac{a_{s}}{a_{dd}} - f\!\left( \sqrt{\lambda}  \right) \right] \, ,
	\label{eq:HRotor011}
\end{equation}
where  $\overline{\omega}$  is the mean trap frequency and   $a_{ho}$
the corresponding harmonic
oscillator length, while $f(\sqrt{\lambda})$ denotes a dimensionless function
with $f(1/2)\approx 0.5$.
The result is derived within first order perturbation theory in the
small parameter $a_{dd} n/a_{ho}$.  The stability against collapse of the quasi-condensate is
well guaranteed for such small particle numbers.

The last remaining parameter is the tunneling energy $t$.
From the critical value $\gamma_{c}$ in Eq.~(\ref{eq:CritGamma}), we find that
the tunneling has to be suppressed by
the factor $t < V n |\chi_{\bf K}|/8 \approx 6.2 {\rm Hz}$.  Note, that
the allowed energies for the hopping term
increases again with the number of particles. In addition, the
superfluid stiffness involves another
factor $n$. As a consequence, the appearance of the superfluid response 
given by the Kosterlitz-Thouless temperature $T_{\rs
KT}  \sim \hbar^2 n_{s}/m$ exhibits the same scaling
as the transition temperature for the solid, i.e.,  $T_{\rs KT} \sim
n^2 V/2$.  It is this scaling of the critical temperatures
for the solid critical temperature as well as the Kosterlitz-Thouless
temperature, which allows us to improve
the experimental parameters by increasing the particle numbers within each well.
The suitable experiments then can be performed by adiabatically
ramping up the optical lattice in a BEC of
chromium atoms at positive s-wave scattering length $a_{s}$ until the
proper filling and hopping energies are
reached. Within this method, temperatures well below the recoil energy
can be reached \cite{RevModPhys.80.885}. Then, lowering
the s-wave scattering length  allows one to pass through the
supersolid instability and the additional formation of the solid
structure.

Discussions with T. Pfau are acknowledged. The work was supported by the 
Deutsche Forschungsgemeinschaft (DFG) within SFB/TRR 21.

\end{document}